\documentclass[final,2p,times,twocolumn]{elsarticle}

\usepackage{amssymb}

\begin{document}

\begin{frontmatter}



\title{Log-periodic oscillations of transverse momentum distributions}


\author[NCBJ]{Grzegorz Wilk}
\ead{wilk@fuw.edu.pl}


\address[NCBJ]{National Centre for Nuclear Research,
        Department of Fundamental Research, Ho\.za 69, 00-681
        Warsaw, Poland}

\author[JKU]{Zbigniew W\l odarczyk}
\ead{zbigniew.wlodarczyk@ujk.edu.pl}

\address[JKU]{Institute of Physics, Jan Kochanowski University,
\'Swi\c{e}tokrzyska 15, 25-406 Kielce, Poland}

\begin{abstract}
Large $p_T$ transverse momentum distributions apparently exhibit
power-like behavior. We argue that, under closer inspection, this
behavior is in fact decorated with some log-periodic oscillations.
Assuming that this is a genuine effect and not an experimental
artefact, it suggests that either the exponent of the power-like
behavior is in reality complex, or that there is a scale parameter
which exhibits specific log-periodic oscillations. This problem is
discussed using Tsallis distribution with scale parameter $T$. At
this stage we consider both possibilities on equal footing.
\end{abstract}


\end{frontmatter}

\begin{figure}[h]
\vspace{-2mm}
\includegraphics[width=7.5cm]{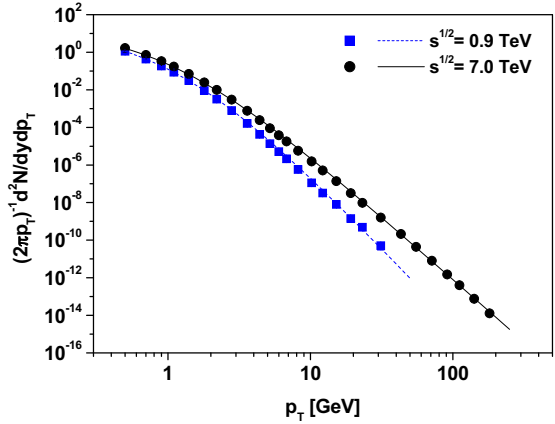}
\vspace{-5mm} \caption{(Color online) Fit to large $p_T$ data for
$pp$ collisions at $0.9$ and $7$ TeV from CMS experiment using
distribution (\ref{eq:Hagedorn}) \cite{CMS}. Parameters used are,
respectively, $(T=0.135,~m=8)$ and $(T=0.145,~m=6.7)$.}
\label{Fig1}
\end{figure}

For some time now it has been popular to fit the different kinds
of transverse momentum spectra measured in multiparticle
production processes to of a Tsallis formula \cite{Tsallis} (cf.,
for example, \cite{qWW,Cleymans,Biro,Deppman,Indian,RW,BYW}). It
can be written in one of two recognized forms: either in original
Tsallis one (with two parameters: $q$ and $T$),
\begin{equation}
f\left( p_T\right) = C\cdot \left[ 1 -
(1-q)\frac{p_T}{T}\right]^{\frac{1}{1-q}} \label{eq:Tsallis}
\end{equation}
or, in the so called "QCD inspired" Hagedorn form
\cite{Hagedorn,H} (with parameters: $m$ and $T$):
\begin{equation}
h\left( p_T\right) = C \left( 1 + \frac{p_T}{mT}\right)^{-m};\quad
m = \frac{1}{q - 1}. \label{eq:Hagedorn}
\end{equation}
For our purposes, these are equivalent (and we shall use them
interchangeably.) They both represent the simplest way of
describing the whole observed range of measured $p_T$
distributions. The best examples are the recent successful fits
\cite{CYWGW} to very large $p_T$ data measured by the LHC
experiments CMS \cite{CMS}, ATLAS \cite{ATLAS} and ALICE
\cite{ALICE} for $pp$ collisions, see Fig. \ref{Fig1}
\footnote{This is the usual domain reserved for the purely
perturbative QCD approach. Nevertheless, it turns out that even
from this kind of approach one can get a distribution with
essentially only two parameters (not counting normalization) which
closely resembles the usual Tsallis distribution
(\ref{eq:Tsallis}), albeit it is not identical with it
\cite{CYWGW1}. Notice that at the midrapidity, i.e., for $y \simeq
0$, and for large transverse momenta, $p_T > M $, one has
$E=\sqrt{M^2 + p^2_T} \cosh(y) \simeq p_T$.}.

\begin{figure}[h]
\vspace{-1mm}
\includegraphics[width=7.5cm]{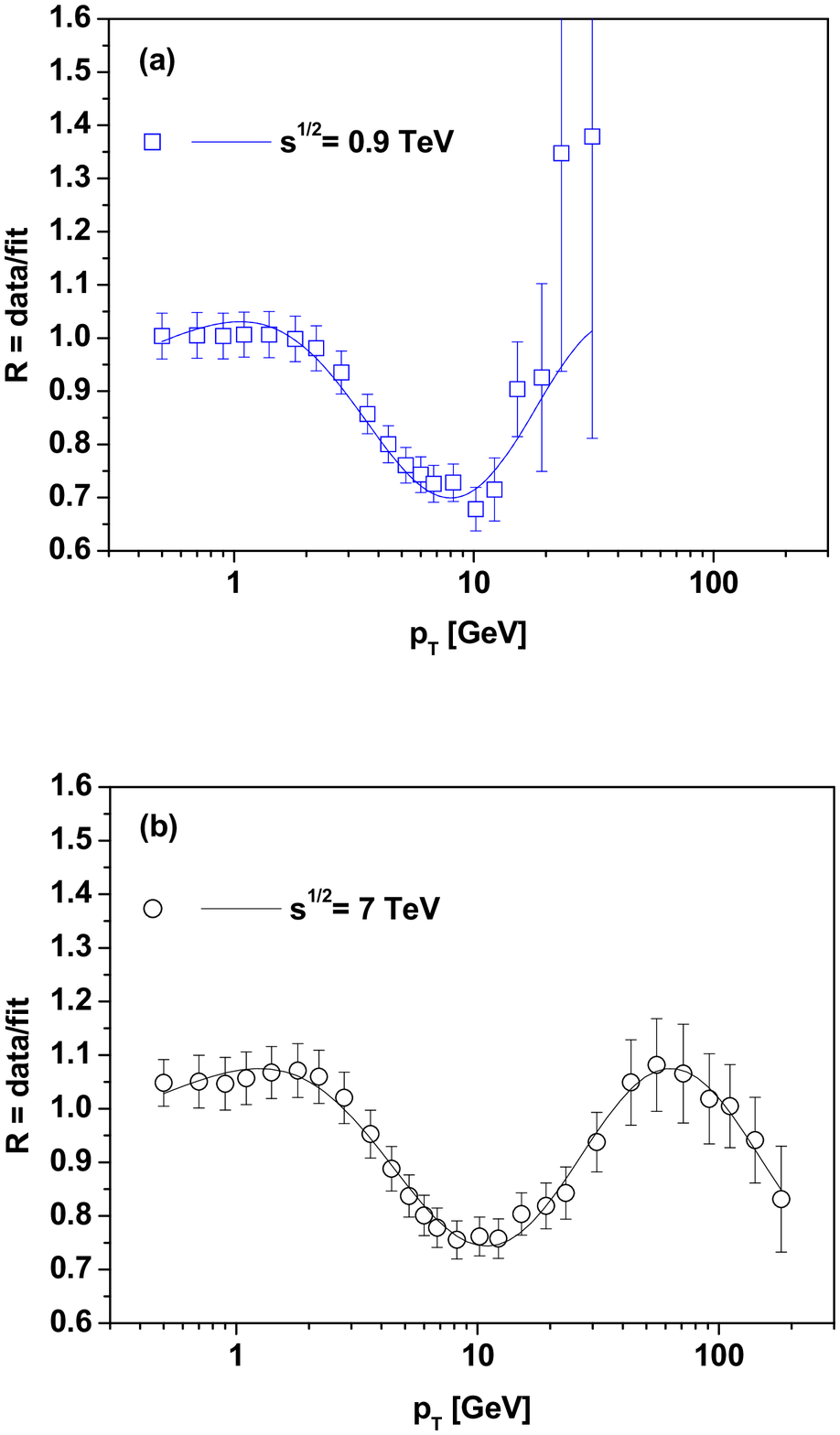}
\vspace{-2mm} \caption{(Color online) Fit to $p_T$ dependence of
data/fit ratio for results from Fig. \ref{Fig1}. Parameters of
function $R$ used (\ref{eq:Factor}) here are, respectively:
$a=0.865$, $c=2.1$ for $0.9$ TeV and $a=0.909$, $c=1.86$ for $7$
TeV, whereas for both energies $b=0.166$, $d=0.948$ and
$f=-1.462$.} \label{Fig2}\vspace{-2mm}
\end{figure}

Albeit both fits look pretty good, closer inspection shows that
ratio of data/fit is not flat but shows some kind of clearly
visible oscillations, cf. Fig. \ref{Fig2}. It turns out that they
cannot be eliminated by suitable changes of parameters $(q,T)$ or
$(m,T)$ in Eqs. (\ref{eq:Tsallis}) or (\ref{eq:Hagedorn}),
respectively\footnote{One has to realize that to really see these
oscillations one needs rather large domain in $p_T$. Therefore,
albeit similar effects can be also seen at lower energies, they
are not so pronounced as here and therefore will not be discussed
at this point.}. In what follows, we assume that this observation
is not an experimental artifact but rather it represents some
genuine dynamical effect which is worth investigating in more
detail.

First notice, that to account for these fluctuations of $f(p_T)$
from Eq. (\ref{eq:Tsallis}) (or $h\left( p_T\right)$ from Eq.
(\ref{eq:Hagedorn})) the original Tsallis formula has to be
multiplied by the following factor (log-oscillating function):
\begin{equation}
R(E)= a + b\cos\left[ c\ln(E + d) + f\right] \label{eq:Factor}
\end{equation}
As recently shown in \cite{cqWW}, such a factor, dressing the
original power law distribution (in our case quasi-power law
Tsallis distribution (\ref{eq:Tsallis})), arises in a natural way
if one allows the power index $q$ to be complex\footnote{There is
vast literature of such situation in different branches of
physics, cf. \cite{Scaling} and other references \cite{cqWW}.}.
For completeness, we shortly explain what this means. In general,
if some function $O(x)$ is scale invariant, i.e., if $O(x) = \mu
O(\lambda x)$, then it must have a power law behavior,
\begin{equation}
O(x) = Cx^{-m}\quad {\rm with}\quad m = \frac{\ln \mu}{\ln
\lambda}. \label{eq:PowerLaw}
\end{equation}
Because one can write $\mu \lambda^{-m} = 1 = e^{i2\pi k}$, where
$k$ is an arbitrary integer, in general,
\begin{equation}
m = - \frac{\ln \mu}{\ln \lambda} + i \frac{2\pi k}{\ln \lambda}.
\label{eq:cq}
\end{equation}
As shown in \cite{cqWW}, the evolution of the differential
$df(E)/dE$ of a Tsallis distribution $f(E)$ with power index $n$
performed for finite differences $dE = \alpha (nT + E)$ (where
$\alpha < nT$ is another new parameter) results in the following
scale invariant relation
\begin{equation}
g[(1 + \alpha )x] = (1 - \alpha n)g(x) \label{eq:scaling}
\end{equation}
where $x = 1 + E/(nT)$. This means that, in general, one can write
Eq. (\ref{eq:Tsallis}) in the form:
\begin{equation}
 g(x) = x^{-m_k},\quad m_k = - \frac{\ln ( 1 - \alpha n)}{\ln (1 +
 \alpha)} + ik \frac{2\pi}{\ln(1 + \alpha)}. \label{eq:solution}
\end{equation}
The power index in Eq. (\ref{eq:solution}) (and in Eq.
(\ref{eq:Tsallis})) is therefore a complex number, the imaginary
part of which signals a hierarchy of scales leading to the
log-periodic oscillations. The meaning of the parameter $\alpha$
becomes clear by noticing that in the special case of $k=0$, for
which one recovers the usual real power law solution, $m_0$
corresponds to fully continuous scale invariance\footnote{In this
case power law exponent $m_0$ still depends on $\alpha$ and
increases with it roughly as $m_0 \simeq n + \frac{n}{2}
(n+1)\alpha + \frac{n}{12}\left(4n^2 + 3n -1\right)\alpha^2 +
\frac{n}{24}\left( 6n^3 + 4n^2 - n +1\right)\alpha^3 + \dots$.
Notice also that $ \alpha < 1/n$. }. In this case one recovers in
the limit $\alpha \rightarrow 0$ the power $n$ in the usual
Tsallis distribution. However, in general one has
\begin{figure}[t]
\vspace{3mm}
\includegraphics[width=7.5cm]{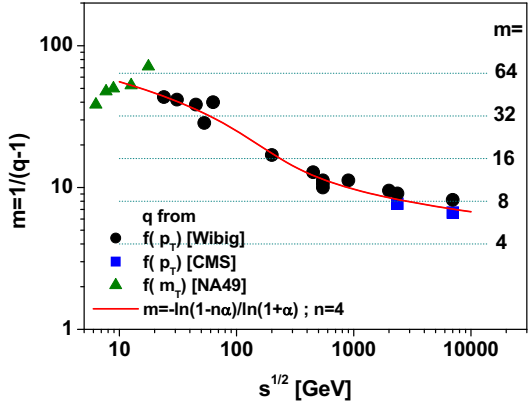}\vspace{-3mm}
\caption{(Color online)  The energy dependence of $m = m_0$
deduced from data \cite{Wibig,CMS,NA49}.} \label{Fig3}
\vspace{-3mm}
\end{figure}
\begin{eqnarray}
g(x) &=& \sum_{k=0}w_k\cdot {\rm Re}\left( x^{-m_k}\right)
=\nonumber\\
&=& x^{- {\rm Re}\left( m_k\right)}\sum_{k=0}w_k\cdot \cos\left[
{\rm Im}\left(m_k\right) \ln(x) \right]. \label{eq:fin}
\end{eqnarray}
This is a general form of a Tsallis distribution for complex
values of the nonextensivity parameter $q$. It consists of the
usual Tsallis form (albeit with a modified power exponent) and a
dressing factor which has the form of a sum of log-oscillating
components, numbered by parameter $k$. Because we do not know {\it
a priori} the details of dynamics of processes under consideration
(i.e., we do not known the weights $w_k$), in what follows we use
only $k=0$ and $k=1$ terms. We obtain approximately,
\begin{equation}
g(E)\! \simeq\! \left( 1\! +\! \frac{E}{nT}\right)^{-m_0}\left\{
w_0 + w_1\cos\left[ \frac{2\pi}{\ln (1\! +\! \alpha)} \ln \left( 1
\!+\! \frac{E}{nT}\right)\right]\right\}. \label{eq:approx}
\end{equation}
In this case one could expect that parameters in general
modulating factor $R$ in Eq. (\ref{eq:Factor}) could be identified
as follows:
\begin{equation}
a = w_0,~b = w_1,~c=\frac{2\pi}{\ln (1 + \alpha)},~d = nT,~f = -
c\cdot \ln(nT). \label{eq:parameters}
\end{equation}
Comparison of the fit parameters of the oscillating term  $R$ in
Eq. ({\ref{eq:Factor}) with Eq. (\ref{eq:solution}) clearly shows
that the observed frequency, here given by the parameter $c$, is
more than an order of magnitude smaller than the expected value
equal to $2\pi/\ln (1 + \alpha)$ for any reasonable value of
$\alpha$. To explain this, notice that in our formalism leading to
Eq. (\ref{eq:approx})) only one evolution step is assumed, whereas
in reality we have a whole hierarchy of $\kappa$ evolutions. This
results (cf. \cite{cqWW}) in the scale parameter $c$ being
$\kappa$ times smaller than in (\ref{eq:approx}),
\begin{equation}
c = \frac{2 \pi}{\kappa \ln (1 + \alpha)}. \label{eq:kappa}
\end{equation}
Experimental data indicate that $\kappa \simeq 22$  (for $\alpha
\simeq 0.15$ and $c \simeq 2$ ).

\begin{figure}[h]
\includegraphics[width=7.5cm]{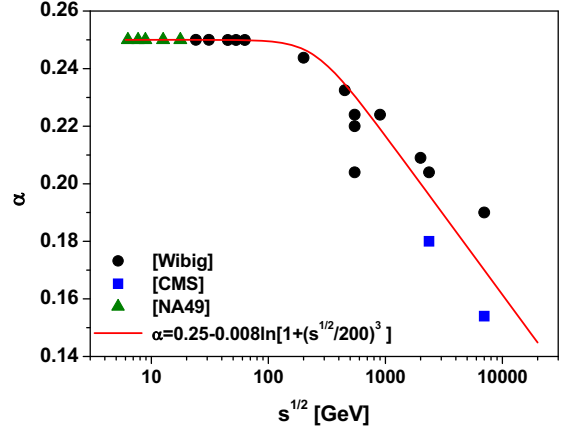} \vspace{-4mm}
\caption{(Color online) The energy dependence of parameter
$\alpha$ present in $m_0$ plotted in Fig. \ref{Fig3}.}
\label{Fig4} \vspace{-1mm}
\end{figure}

From Eq.(\ref{eq:solution}) we see that $m_0 > n$. This suggests
the following explanation of the difference seen between
prediction from theory and the experimental data: the measurements
in which log-periodic oscillations appear underestimate the true
value that follows from the underlying dynamics which leads to the
smooth Tsallis distribution. As an example consider the $m_0$
dependence on $\alpha$, assuming the initial slope $n = 4$ (this
is the value of $n$ expected from the pure QCD considerations for
partonic interactions \cite{CYWGW1}). The energy behavior of the
power index $m_0$ in the Tsallis part is shown in Fig. \ref{Fig3},
whereas the energy dependence of the parameter $\alpha$ contained
in $m_0$ is shown in Fig. \ref{Fig4}.

So far we attributed the observed log-periodic oscillations to the
complex values of the power index $m$ (i.e., to the complex
nonextensivity parameter $q$)\footnote{The possible dynamical
implication of this fact, cf. for example \cite{Scaling,cqWW} and
remarks in footnote 7, is outside of the scope of present paper.}.
However, this phenomenon can be also explained in a completely
different way, namely by keeping the nonextensivity parameter $q$
real (as in the original Tsallis distribution) but instead
allowing the parameter $T$ to oscillate in a specific way as
displayed in Fig. \ref{Fig5}. As seen there, the observed
log-periodic oscillations of $R$ can be reproduced by a suitable
$p_T$ dependence of the scale parameter (the temperature) $T$,
present in Tsallis distribution, here expressed by following a
general formula (resembling Eq. (\ref{eq:Factor}), with generally
energy dependent fit parameters $(\bar{a}, \bar{b}, \bar{c},
\bar{d}, \bar{f})$):
\begin{equation}
T = \bar{a} + \bar{b}\sin\left[ \bar{c} \left( \ln(E +
\bar{d}\right) + \bar{f} \right] \label{eq:T}
\end{equation}
\begin{figure}[h]
\vspace{-5mm}
\includegraphics[width=8cm]{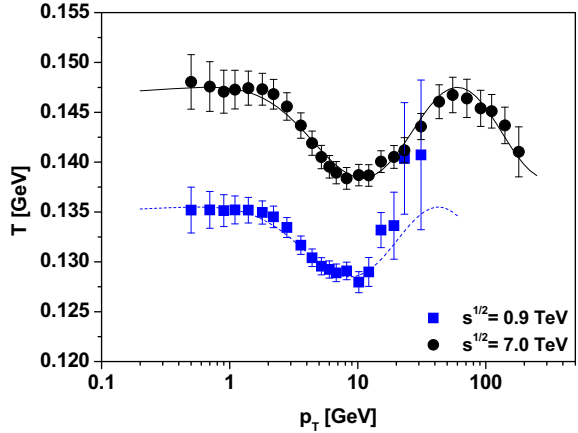}
\vspace{-7mm} \caption{(Color online) The $T=T\left( p_T\right)$
for Eq. (\ref{eq:T}) for which $R=1$. Parameters used are:
$\bar{a} = 0.132,~\bar{b} = 0.0035,~\bar{c} = 2.2,~\bar{d} =
2.0,~\bar{f} = -0.5$ for $0.9$ TeV and $\bar{a} = 0.143,~\bar{b} =
0.0045,~\bar{c} = 2.0,~\bar{d} = 2.0,~\bar{f} = -0.4$ for $7$
TeV.} \label{Fig5}
\end{figure}

To explain such behavior, start with the well known
\cite{WellKnown} stochastic equation for the temperature
evolution, which in the Langevin formulation has the form (in
which we allow for an energy dependent noise, $\xi(t,E)$):
\begin{equation}
\frac{dT}{dt} + \frac{1}{\tau} T + \xi(t,E) T = \Phi.
\label{eq:St1}
\end{equation}
For the time dependent $E=E(t)$ it reads:
\begin{equation}
\frac{dT}{dE}\frac{dE}{dt} + \frac{1}{\tau} T + \xi(t,E) T = \Phi.
\label{eq:St2}
\end{equation}
In the scenario of {\it preferential attachment} (known from the
growth of networks \cite{NETWORKS}) one expects
that\footnote{Notice that in the usually used multiplicative noise
scenario described by $\gamma(t)$, not discussed here, one has
$\frac{dE}{dt} = \gamma(t) E + \xi (t)$.}
\begin{equation}
\frac{dE}{dt} = \frac{E}{n} + T. \label{eq:St3}
\end{equation}
Using now Eq. (\ref{eq:St3}) one can write Eq. (\ref{eq:St2} as
\begin{equation}
\left(\frac{E}{n} + T\right)\frac{dT}{dE} + \frac{1}{\tau} T +
\xi(t,E) T = \Phi .\label{eq:St4}
\end{equation}
This can be subsequently transformed to
\begin{equation}
\left( \frac{1}{n} + T e^{-\ln E}\right)\frac{dT}{d(\ln E)} +
\frac{1}{\tau} T + \xi(t,E) T = \Phi \label{eq:eq:St4a}
\end{equation}
and, after differentiating, to
\begin{eqnarray}
&&\left( \frac{1}{n} + T e^{-\ln E}\right)\frac{d^2T}{d(\ln E)^2}
+ \left[\frac{dT}{d(\ln E)}\right]^2 e^{-\ln E}
-\nonumber\\
&& - \left[  T e^{-\ln E} - \frac{1}{\tau} -
\xi(t,E)\right]\frac{dT}{d(\ln E)}
+\nonumber\\
&& + T\frac{d\xi(t,E)}{d(\ln E)} = 0. \label{eq:St4b}
\end{eqnarray}
For large $E$ (i.e., neglecting terms $\propto 1/E$) one obtains
the following equation for $T$:
\begin{equation}
\frac{1}{n}\frac{d^2T}{d(\ln E)^2} + \left[ \frac{1}{\tau} +
\xi(t,E) \right]\frac{dT}{d(\ln E)} + T\frac{d\xi(t,E)}{d\ln E)} =
0.\label{eq:St5}
\end{equation}
Now assume that noise $\xi(t,E)$ increases logarithmically with
energy,
\begin{equation}
\xi(t,E) = \xi_0(t) + \frac{\omega^2}{n}\ln E. \label{eq:St6}
\end{equation}
For this choice of noise Eq. (\ref{eq:St5}) is just an equation
for the damped hadronic oscillator and has a solution in the form
of log-periodic oscillation of temperature with frequency
$\omega$:
\begin{equation}
T = C \exp\left\{ - n\cdot\left[\frac{1}{2\tau} +
\frac{\xi(t,E)}{2}\right]\ln E\right\}\cdot \sin(\omega\ln E +
\phi). \label{eq:St7}
\end{equation}
The phase shift parameter $\phi$ depends on the unknown initial
conditions and is therefore an additional fitting parameter.
Averaging the noise fluctuations over time $t$ and taking into
account that the  noise term cannot on average change the
temperature (cf. Eq. (\ref{eq:St1}) in which $\langle dT/dt\rangle
= 0$ for $\Phi = 0$), i.e., that
\begin{equation}
\frac{1}{\tau} + \langle \xi(t,E)\rangle = 0, \label{eq:St8}
\end{equation}
we have
\begin{equation}
T = \bar{a} + \frac{b'}{n}\sin( \omega \ln E + \phi).
\label{eq:St9}
\end{equation}
The amplitude of oscillations, $b'/n$, comes from the assumed
behavior of the noise as given in Eq. (\ref{eq:St6}). Notice that
for large $n$, the energy dependence of the noise disappears (and
because, in general, $n$ decreases with energy, one can therefore
expect only negligible oscillations for lower energies but
increasing with the energy).  This should now be compared with the
parametrization of $T(E)$ given by Eq. (\ref{eq:T}) and used to
fit data in Fig.\ref{Fig5}. Looking at parameters we can see that
only a small amount of $T$ (of the order of $\bar{b}/\bar{a} \sim
3\%$) comes from the stochastic process with energy dependent
noise, whereas the main contribution emerges from the usual
energy-independent Gaussian white noise.

To conclude, the above oscillating $T$ needed to fit the
log-periodic oscillations seen in data can be obtained in yet
another way. So far we were assuming that the noise $\xi(t,E)$ has
the form of Eq. (\ref{eq:St6}) and, at the same time, we were
keeping the relaxation time $\tau$ constant. However, it turns out
that we could equivalently assume the energy $E$ independent white
noise, $\xi(t,E) = \xi_0(t)$, but allow for the energy dependent
relaxation time taken in the form of
\begin{equation}
\tau = \tau(E)  = \frac{n\tau_0}{n + \omega^2 \ln E}.
\label{ew:tauE}
\end{equation}
This assumption corresponds to the following time evolution of the
temperature,
\begin{equation}
T(t) = \langle T\rangle + [T(t=0) - \langle T\rangle]
E^{-t\omega^2/n} \exp\left(-\frac{t}{\tau_0}\right),
\label{eq:Ttau}
\end{equation}
which is gradually approaching its equilibrium value $\langle T
\rangle$ and reaches it more quickly for higher energies.

To summarize, we have presented two possible mechanisms which
could result in the log-periodic oscillations apparently present
in data for transverse momentum distributions observed in LHC
experiments. In both cases one uses a Tsallis formula (either in
the form of Eq. (\ref{eq:Tsallis}) or Eq. (\ref{eq:Hagedorn})),
with main parameters $m$ - the scaling power exponent (or
nonextensivity $q = 1 + 1/m$) and $T$ - the scale parameter
(temperature). In the first approach, our Tsallis distribution is
decorated by the oscillating factor which emerges in a natural way
in the case of complex power exponent $m$ (or complex
nonextensivity $q$)\footnote{It is worth to mention at this point
that complex $q$ inevitably means also complex heat capacity $C =
1/(1 - q)$ (c.f., \cite{qWW,BiroC} and also \cite{Campisi}). Such
complex (frequency dependent) heat capacities are widely known and
investigated, see \cite{HC}.} with the scale parameter $T$
remaining untouched. In the second approach, it is the other way
around, i.e., whereas $m$ (or $n$ as in Eq. (\ref{eq:St7}))
remains untouched, the scale parameter $T$ is now oscillating.
From Eq. (\ref{eq:St9}) one can see that $T=T(n=1/(1-q),E)$ and as
a function of nonextensivity parameter $q$ it continues our
previous efforts to introduce an effective temperature into the
Tsallis distribution, $T_{eff} = T(q)$, here in a much more
general form as in \cite{qWW} or \cite{RODOS}. The two possible
mechanisms resulting in such $T$ were outlined: the energy
dependent noise connected with the constant relaxation time, or
else the energy independent white noise, but with energy dependent
relaxation time.

We close by noting that, at the present level of investigation, we
are not able to indicate which of the two possible mechanisms
presented here (complex $q$ or oscillating $T$) and resulting in
log-periodic oscillations is the preferred one. This would demand
more detailed studies on the possible connections with dynamical
pictures. For example, as discussed long time ago by studying
apparently similar effects in some exclusive reactions using the
QCD Coulomb phase shift idea \cite{Pire}. The occurrence of some
kind of complex power exponents was noticed there as well, albeit
on completely different grounds than in our case. A possible link
with our present analysis would be very interesting but would
demand an involved and thorough analysis.

\vspace*{0.3cm} \noindent {\bf Acknowledgments}

This research  was supported in part by the National Science
Center (NCN) under contract Nr 2013/08/M/ST2/00598. We would like
to warmly thank Dr Eryk Infeld for reading this manuscript.


\begin{thebibliography}{99}

\bibitem{Tsallis} C.~Tsallis, J.\ Statist.\ Phys.\ {\bf 52} (1988) 479 (1988);
                  Eur.\ Phys.\ J.\ A {\bf 40} (2009) 257 (2009)
                  and {\it Introduction to Nonextensive Statistical Mechanics}
                  (Springer, 2009). For an updated bibliography on this subject,
                  see http://tsallis.cat.cbpf.br/biblio. htm.

\bibitem{qWW} G.~Wilk and Z.~W\l odarczyk,  Eur.\ Phys.\ J.\ A {\bf 40}
              (2009) 299; {\bf 48} (2012) 161; Cent.\ Eur.\ J.\ Phys.\ {\bf 10} (2012) 568.

\bibitem{Cleymans} J.~Cleymans and D.~Worku, J.\ Phys.\ G {\bf39} (2012) 025006;
                   Eur.\ Phys.\ J.\ A {\bf 48} (2012) 160.

\bibitem{Biro} K.~\"Urm\"osy, G.~G.~Barnaf\"oldi and T.~S.~Bir\'o,
               Phys.\ Lett.\ B\ {\bf 701} (2012) 111 and {\bf 718} (2012) 125.

\bibitem{Deppman} I.~Sena and A.~Deppman, Eur.\ Phys.\ J.\ A {\bf 49} (2013) 17.

\bibitem{Indian} P.~K.~Khandai, P.~Sett, P.~Shukla and V.~Singh, Int.\ J.\ Mod.\
                 Phys.\ A {\bf 28} (2013) 1350066 and J.\ Phys.\ G {\bf 41} (2014) 025105.

\bibitem{RW} M.~Rybczy\'nski and Z.~W\l odarczyk, Eur.\ Phys.\ J.\ C {\bf
             74}(2014)2785.

\bibitem{BYW} Bao-Chun~Li.,Ya-Zhou~Wang and Fu-Hu~Liu, Phys.\
              Lett.\ B {\bf 725} (2013) 352.

\bibitem{Hagedorn} C.~Michael and L.~Vanryckeghem, J.\ Phys.\ G {\bf
                   3} (1977) L151; C.~Michael, Prog.\ Part.\ Nucl.\ Phys.\
                   {\bf 2} (1979) 1.

\bibitem{H} G.~Arnison et al (UA1 Collab.), Phys.\ Lett.\ B {\bf 118}  (1982) 167;
            R.~Hagedorn, Riv.\ Nuovo\ Cimento\ {\bf 6} (1984) 1.

\bibitem{CYWGW} C-Y.~Wong and G.~Wilk, Acta\ Phys.\ Polon. B {\bf
                43} (2012) 2047.

\bibitem{CMS} V.~Khachatryan $et~al.$ (CMS Collaboration), JHEP\ {\bf 02} (2010) 041
              and JHEP\ {\bf 08} (2011) 086; Phys.\ Rev.\ Lett.\  {\bf 105} (2010) 022002.

\bibitem{ATLAS} G.~Aad $et~al.$ (ATLAS Collaboration), New\ J.\ Phys.\ {\bf 3} (2011) 053033.

\bibitem{ALICE} B. Abelev et al. (ALICE Collaboration), Phys. Lett. B {\bf 722} (2013) 262.

\bibitem{CYWGW1} C-Y.~Wong and G.~Wilk, Phys.\ Rev.\ D\ {\bf 87} (2013)
                 114007 and {\it Relativistic Hard-Scattering and Tsallis Fits to
                 $p_T$ Spectra in $pp$ Collisions at the LHC}, arXiv:1309.7330[hep-ph],
                 to be published in The\ Open\ Nuclear\ \& Particle\ Physics\ Journal\
                 (2014).

\bibitem{cqWW}  G.~Wilk and Z.~W\l odarczyk, {\it Tsallis distribution with complex nonextensivity
                parameter $q$}, arXiv:1403.3263 [cond-mat.stat-mech].

\bibitem{Scaling} D.~ Sornette, Phys.\ Rep.\ {\bf 297} (1998) 239.

\bibitem{WellKnown}   N.~G.~van Kampen, {\it Stochastic Processes in Physics and Chemistry}
                     (Elsevier Science Publishers B.V., North-Holland, Amsterdam, 1987),
                     Chap. VIII.

\bibitem{NETWORKS}  G.~Wilk and Z.~W\l odarczyk, Acta\ Phys.\ Polon.\ B {\bf 35} (2004) 871
                    and B {\bf 36} (2005) 2513; Eur.\ Phys.\ J.\ A {\bf 48} (2012) 162.

\bibitem{Wibig} T.~ Wibig, J.\ Phys.\ G {\bf37} (2010) 115009.

\bibitem{NA49} C.~ Alt et al., Phys.\ Rev.\ C {\bf77} (2008) 034906 and C {\bf 77} (2008)
               024903; S.~ V.~ Afanasiev et al., Phys.\ Rev.\ C {\bf 66} (2002) 054902.

\bibitem{BiroC} T.~S.~ Bir\'o, G.~G.~Barnaf\"oldi and P.~V\'an, Eur.\ Phys.\ J.\ A {\bf 49}
                (2013) 110.

\bibitem{Campisi} M.~Campisi, Phys.\ Lett.\ A {\bf 366} (2007)
                  335; A~.R.~Plastino and A.~Plastino, Phys.\ Lett.\ A {\bf 193} (1994) 140;
                  M.~P.~Almeida, Physica\ A {\bf 300} (2001) 424.

\bibitem{HC} J.~E.~K.~Schawe, Thermochim.\ Acta\ {\bf 260} (1995) 1; J~.-L.~Garden,
             Thermochimica\ Acta\ {\bf 460} (2007) 85.


\bibitem{RODOS} G.~Wilk and Z.~W\l odarczyk; AIP\ Conf.\ Proc.\ {\bf 1558} (2013)
                893; arXiv:1307.7855.


\bibitem{Pire} B.~Pire and J.~P.~Ralston, Phys.\ Lett.\ B {\bf
               117} (1982) 233; J.~P.~Ralston and B.~Pire, Phys.\
               Rev.\ Lett.\ {\bf 49} (1982) 1605 and {\bf 61}
               (1988) 1823.


\end{thebibliography}
\end{document}